\documentclass[twocolumn,prl,tightenlines,nofootinbib,amsmath,amssymb,amsfonts,superscriptaddress]{revtex4}
\usepackage[pass,letterpaper]{geometry}

\usepackage{aas_macros}
\usepackage{hyperref}
\usepackage{lipsum}
\usepackage{graphicx}

\usepackage{xcolor}
\usepackage[normalem]{ulem}

\newcommand{\be}{\begin{equation}}
\newcommand{\ee}{\end{equation}}
\newcommand{\bea}{\begin{eqnarray}}
\newcommand{\eea}{\end{eqnarray}}
\newcommand{\dd}{{\rm d}}
\usepackage{bm}

\newcommand{\gr}[1]{{\bm #1}}

\newcommand{\com}[1]{}

\newcommand{\FPM}{{\texttt{FPM}}}
\newcommand{\SIT}{{\texttt{SITS}}}

\begin{document}

\title{A precise numerical estimation of the magnetic field generated
  around recombination}

\author{Christian Fidler}
\email{christian.fidler@uclouvain.be }
\affiliation{Catholic University of Louvain - Center for Cosmology, Particle Physics and Phenomenology (CP3)
2, Chemin du Cyclotron, B-1348 Louvain-la-Neuve, Belgium}

\author{Guido Pettinari}
\email{guido.pettinari@gmail.com}
\noaffiliation

\author{Cyril Pitrou}
\email{pitrou@iap.fr}
\affiliation{Institut d'Astrophysique de Paris, CNRS-UMR 7095, Universit\'e Pierre~\&~Marie Curie - Paris VI, Sorbonne
  Universit\'es, 98 bis Bd Arago, 75014 Paris, France}

\date{\today}


\begin{abstract}
We investigate the generation of magnetic fields from non-linear effects around recombination. As tight-coupling is gradually lost when approaching $z\simeq 1100$, the velocity difference between photons and baryons starts to increase, leading to an increasing Compton drag of the photons on the electrons. The protons are then forced to follow the electrons due to the electric field created by the charge displacement; the same field, following Maxwell's laws, eventually induces a magnetic field on cosmological scales. Since scalar perturbations do not generate any magnetic field as they are curl-free, one has to resort to second-order perturbation theory to compute the magnetic field generated by this effect. We reinvestigate this problem numerically using the powerful second-order Boltzmann code \textsf{SONG}. We show that: {\it i)} all previous studies do not have a high enough angular resolution to reach a precise and consistent estimation of the magnetic field spectrum; {\it ii)} the magnetic field is generated up to $z\simeq 10$; {\it iii)} it is in practice impossible to compute the magnetic field with a Boltzmann code for scales smaller than $1\,{\rm Mpc}$. Finally we confirm that for scales of a few ${\rm Mpc}$, this magnetic field is of order $2\times 10^{-29}{\rm G}$, many orders of magnitude smaller than what is currently observed on intergalactic scales.
\end{abstract}

\maketitle



\section*{Introduction}

As magnetic fields have been firmly detected in galaxies and galaxy clusters, evidence is growing for magnetic fields in the intergalactic medium (see e.g. the reviews \cite{Ryu2011,Widrow2011,Durrer2013}). Lower bounds of the order $10^{-16} {\rm G}$ have been reported (see e.g. \cite{Neronov2010,Essey2011,Takahashi2013}) on scales of order $10 {\rm Mpc}$ corresponding to the typical size of voids in the cosmic web, using the delayed secondary emission of cosmic rays. 
Conversely, upper bounds of the order of a few ${\rm nG}$ have been placed on the strength of the magnetic field in the intergalactic medium, using the Faraday rotation that such fields would induce on the cosmic microwave background polarization~\cite{Planck2015Faraday,PolarBear}. It is all the more important to understand the origin of intergalactic magnetic fields as they are thought to be the seed fields at the origin of the stronger magnetic fields inside galaxies and clusters. 

Several mechanisms have been proposed to explain the origin of these seed fields. Among them, non-conformal couplings between the inflaton and the electromagnetic field during inflation have been investigated (see e.g.~\cite{Martin:2007ue,Fujita:2015iga,Subramanian2010}) as they can generate large coherence scales. However there are strong constraints on these models since after being generated, all cosmological magnetic fields decay adiabatically with the expansion as $1/a^2$. One may thus wonder if the large-scale seed magnetic fields could be generated much later after the end of inflation, in the primordial plasma. 

Indeed, it has been shown~\cite{Ichiki2006,Ichiki2007,Maeda2008,Fenu2010} that vortical currents create magnetic fields around recombination. As tight-coupling between photons and baryons is gradually lost (see e.g. \cite{Pitrou-TC-2011,CyrRacine}), the photons drag the electrons through Compton interactions thus generating a vortical electric field which eventually forces the protons to move with the electrons. By means of Maxwell's laws of electromagnetism, the electric field sources a magnetic field which, being generated around recombination ($z\sim1100$), is present today diminished by an adiabatic factor of just $\sim10^{-6}$.

The vortical currents required to form the magnetic field cannot be generated at first order in the cosmological perturbations, due to the suppression of vector and tensor modes. At second order, however, vorticity arises naturally and eventually leads to a rather small intergalactic magnetic fields, typically a few $10^{-29} {\rm G}$ for scales of order $10 {\rm Mpc}$ today. Nevertheless, this intrinsic magnetic field is unavoidable as it is a prediction of the standard cosmological model not involving speculative physics.

\begin{figure}[!htb]
	\includegraphics[width=8.5cm]{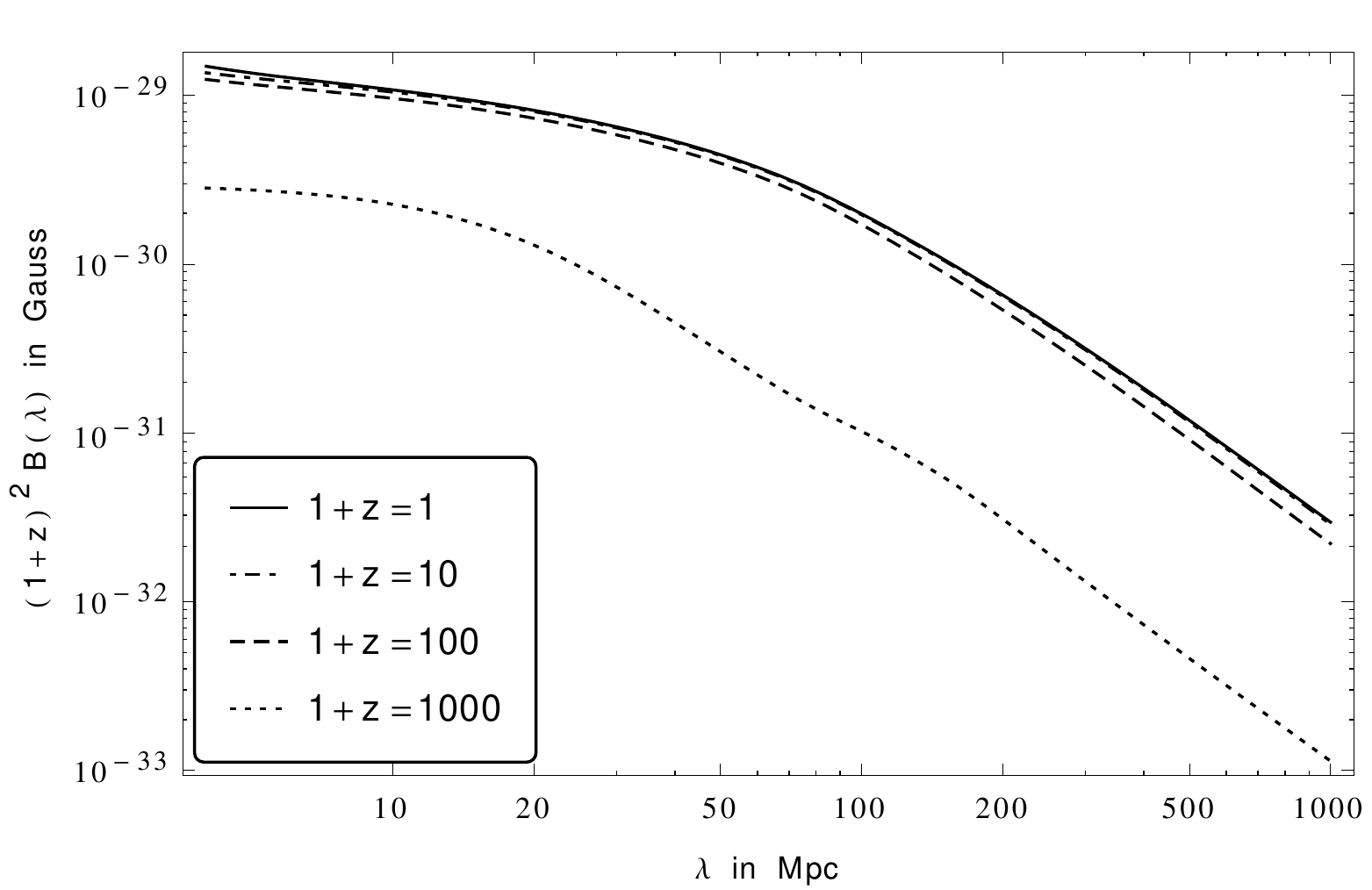}
	\caption{The averaged magnetic field $B(\lambda)$ as a function of the averaging scale for various redshifts. We apply a factor $(1+z)^2$ to cancel the adiabatic decay of the magnetic field. The curves at $1+z = 10$ and $1+z = 1$ overlap since there are no sources for the magnetic field at late times. In contrast, there is a significant difference between the magnetic field after recombination at $1+z = 1000$ and $1+z = 100$ demonstrating the continuation of the sources after recombination.}
	\label{fig1}
\end{figure}

There is an ongoing debate on the asymptotic behaviour of this intrinsic magnetic field. In this paper we compute it with the greatest accuracy so far, using the state of the art second-order Boltzmann code \textsf{SONG}. In Fig.~\ref{fig1}, we show our most important result, the intrinsic magnetic field as a function of the smoothing scale.

At early times, we recover the analytic limit $\sqrt{k^3 P_B(k)}\propto k^{7/2}$ found numerically for large scales in Refs. \cite{Ichiki2007,Saga2015}, hereafter called \SIT, and we are able to find an analytic explanation for this limit. This contrasts with the limit $\sqrt{k^3 P_B(k)}\propto k^4$ derived in Ref. \cite{Fenu2010}, hereafter called \FPM, or Ref. \cite{Nalson2013} where the sub-Hubble modes were inconsistently ignored.
From recombination onwards, we essentially recover the results of \FPM~ for the shape and magnitude of the magnetic field spectrum.
We find however that in order to obtain a better precision up to a scale of a few ${\rm Mpc}$, we should increase the angular accuracy of the Boltzmann code by computing photon multipoles up to $\ell_{\rm max}=100$, since the magnetic field keeps being generated up to $z\simeq 10$.
In addition the magnetic field sources at these times are subject to important higher order corrections on small scales. 
Hence, we find that it is impractical to push the numerical computation in a Boltzmann code beyond scales of order $1{\rm Mpc}$ as done in \SIT.

\section{Magnetic field generation in the primordial plasma}

Before recombination, electrons and photons are tightly coupled by Compton scattering. Due to the overabundance of photons, the electrons follow the fluid of photons. On the other hand, the much heavier protons are almost unaffected by the photons. The Compton scattering thus creates a tension between the electrons and protons which sources an electric field that counters the displacement of charges. When comparing the various time-scales of the problem, it is found that the protons follow the electrons nearly instantly, and together they can thus be considered as a fluid of {\it baryons}. The resulting electric field $E^\mu$ seen by a cosmological observer with four-velocity $u^\mu$ is given by~\cite{Ichiki2007,Fenu2010,Nalson2013}
\be\label{GenerationE}
e n_e E^\mu = h^\mu_\nu \nabla_\alpha T_{\rm b}^{\alpha\nu} =
h^\mu_\nu C^\nu_{\gamma \to {\rm b}}\,,\quad
h^\mu_\nu \equiv \delta^\mu_\nu+u^\mu u_\nu\,,
\ee 
where $n_e$ is the number of free electrons, $T_{\rm b}^{\mu\nu}$ the stress-energy tensor of baryons and $C_{\gamma \to {\rm b}}^\mu$ the collision term induced by the Compton interaction of photons on baryons. The evolution of the magnetic field is then inferred from the Maxwell equations, more precisely from the structure equation
\be\label{MaxwellB}
\nabla_{[\alpha} F_{\mu\nu]} = 0\,.
\ee

In order to obtain quantitative results for the primordial magnetic field, Eqs. \eqref{GenerationE} and \eqref{MaxwellB} must be expressed in a cosmological context. To this end we consider a homogeneous and isotropic background, on which we add scalar perturbations. The corresponding metric is
\be\label{Metric}
\dd s^2= a(\eta)^2 \left[ -(1+2\Phi)\dd \eta^2 + (1-2\Psi)\delta_{ij} \dd x^i \dd x^j\right]\,.
\ee
where $\eta$ is the conformal time, $a(\eta)$ is the scale factor which encodes the expansion of the universe, and $\Phi$ and $\Psi$ are two gravitational potentials which are nearly equal for most of the cosmological history and are identified with the Newtonian potential on small scales.

Since the electric and magnetic field are frame dependent, we need to specify with respect to which observer they are defined. In the cosmological context, it is natural to define these fields with respect to an observer whose velocity is given by $u_\mu \propto \dd \eta_\mu$, and we are eventually interested in the time evolution of the magnetic field seen by these cosmological observers. From \eqref{MaxwellB}, the time evolution of the intrinsic magnetic field is simply given by
\be
\label{eq:Bevolution}
\frac{\partial (a^2 B^i)}{\partial \eta} = -a^2 \varepsilon^{ijk} \partial_j[(1+ \Phi-\Psi)E_k]\,.
\ee
In this expression, all indices except the one on the derivative are taken in a local orthonormal frame and $\varepsilon^{ijk}$ is the antisymmetric tensor with $\varepsilon^{123}=1$.

A first consequence is that we need to compute the electric field up to second-order in cosmological perturbations. Indeed at first order, since we have only scalar perturbations, $E_i \propto \partial_i \Phi$ which from \eqref{eq:Bevolution} implies that no magnetic field is generated and it simply decays adiabatically as $1/a^2$.

\section{Power spectra and averaged magnetic field}

The full set of second-order equations including all types of matter (cold dark matter, baryons, photons, neutrinos) with the perturbations of the metric up to second order can be found in \cite{Nakamura2006,Pitrou2008,Beneke2010,Naruko2013}. We need to integrate these equations numerically to obtain the sources \eqref{GenerationE} of the electric field, including all second-order effects. In order to integrate ordinary differential equations, instead of partial differential equations, the calculation is performed in Fourier space, and the contributions of the various Fourier modes are summed at the end. Since relativistic particles cannot be consistently described by a perfect fluid, photons and neutrinos are described statistically and their evolution is given by a Boltzmann equation. The codes evolving the perturbations of the metric and of the fluids in Fourier space are thus called Einstein-Boltzmann codes and we base our numerical analysis on \textsf{ SONG}~\cite{SONG1,SONGB,SONGPolar}.

In practice, it proves easier to decompose the magnetic field on two polarization vectors $\gr{e}_\pm$  which are orthogonal to  the direction $\hat{\gr{k}}$ of the Fourier mode considered, given that the magnetic field is divergenceless. We thus use that in Fourier space $\gr{B} = B^+ \gr{e}_+ + B^- \gr{e}_-$. The result of the second-order numerical integration is given by transfer functions ${\cal T}^\pm$ which are defined as the convolution kernels of the magnetic field:
\be\label{DefBeta}
B^\pm(\gr{k},\eta)= \int \frac{\dd^3 \gr{q}}{(2 \pi)^3}{\cal T}^\pm_B(\gr{q},\gr{k}-\gr{q},\eta) \Phi_{\rm in}(\gr{k}) \Phi_{\rm in}(\gr{k}-\gr{q}) \,.
\ee
Note that the ${\cal T}^\pm(\gr{k}_1,\gr{k}_2,\eta)$ can be chosen to be symmetric in $\gr{k}_1$ and $\gr{k}_2$ as the antisymmetric part does not contribute to $B^\pm(\gr{k})$ and hereafter we choose to work with such symmetric transfer functions.

The statistical properties of the magnetic field are then inferred from the non-statistical transfer functions and from the statistical properties of the initial gravitational potential in Fourier space $\Phi_{\rm in}(\gr{k})$. For statistically homogeneous and isotropic initial conditions, the two-point correlation function in Fourier space of the initial potential is of the form
\be\label{DefPk}
\langle \Phi_{\rm in}(\gr{k}) \Phi^\star_{\rm in}(\gr{k}')\rangle = (2\pi)^3\delta(\gr{k}-\gr{k}')P(k)\,,
\ee
where $P(k)$ is by definition the initial power spectrum.
The two-point correlation of the magnetic field in Fourier space is immediately deduced from Eqs. \eqref{DefBeta} and \eqref{DefPk}. It is of the form
\be\label{DefPBk}
\langle B(\gr{k},\eta) B^\star(\gr{k}',\eta) \rangle = (2\pi)^3 \delta(\gr{k}-\gr{k}')P_B(k,\eta)\,,
\ee
where $P_B(k,\eta)$ is the magnetic field power spectrum at a given time $\eta$, and it is given by 
\be
\label{eq:PowerSpectrum}
P_B(k,\eta) = 4 \int \frac{\dd^3 \gr{q}}{(2\pi)^3} \left|{\cal T}^+_B(\gr{q},\gr{k}-\gr{q},\eta)\right|^2P(q)P(|\gr{k}-\gr{q}|)\,.
\ee

Eventually we are interested in the shape and magnitude of the magnetic field spectrum today or more generally at low redshift. However, since measurements of the intergalactic magnetic field are made in terms of its magnitude in real space, we need to find a way to relate the Fourier power spectrum to the real space observations. 
Hence, we define the averaged magnetic field over a scale $\lambda$ by the convolution of the magnetic field with a sphere of radius $\lambda$:
\be\label{DefBlambda}
B_\lambda(\gr{x}) \equiv \left(\frac{4}{3}\pi \lambda^3\right)^{-1}\int B(\gr{x}+\gr{y}) \theta(\lambda- |\gr{y}|) \dd^3 \gr{y}\,, 
\ee
where $\theta$ is the Heaviside step function. From \eqref{DefPBk} and \eqref{DefBlambda} we find that the average fluctuations of the smoothed magnetic field are given by
\bea\label{DefBlambdaSquare}
{\cal B}_\lambda^2(\eta) &\equiv& \langle B_\lambda(\gr{x},\eta) B_\lambda(\gr{x},\eta) \rangle \nonumber\\
&=&\frac{1}{2\pi^2}\int k^3 P_B(k,\eta) W^2(k \lambda) \dd \ln k \,.
\eea
with $W(x)=3 j_1(x)/x$. The smoothing scale $\lambda$ introduces a cut-off scale as the contributions of modes with $k \lambda \gg 1$ are effectively removed by the window function $W(x)$ which is unity when $x \to 0$ and vanishes when $x \to \infty$. The definition \eqref{DefBlambda} and its counterpart in Fourier space \eqref{DefBlambdaSquare} are similar to the standard definitions used for the smoothed density field of matter (see e.g. \cite{Peebles1994} for the definition of $\sigma_8$), but note that they are different from the smoothing definitions of Ref.~\cite{Fenu2010}. In Eq.~\eqref{DefBlambdaSquare} we notice that
\be
Q(k,\eta) \equiv \sqrt{k^3 P_B(k,\eta)/(2 \pi^2)}
\ee
characterizes the contribution of a given Fourier mode to the averaged magnetic field ${\cal B}_\lambda$. When representing magnetic field power spectra, we thus plot $Q(k,\eta)$.

\section{Numerical results from \textsf{SONG}}

We compute the magnetic field using the second-order Einstein-Boltzmann code \textsf{SONG}\footnote{A pre-release version is freely available at \href{https://github.com/coccoinomane/song}{https://github.com/coccoinomane/song}.}, which was originally designed for the numerical computation of the intrinsic bispectrum and intrinsic B-mode polarisation of the cosmic microwave background \cite{SONG1,SONGB,SONGPolar}. \textsf{SONG} has been intensely tested by performing various consistency checks, numerical convergence runs, the matching of several analytic limits and most importantly by direct comparison to independent state of the art second-order codes \cite{GuidoThesis,Zhiqi,Su}.

The mechanism generating the intrinsic magnetic field from non-linear perturbations is mainly efficient around the epoch of recombination but, as opposed to the cosmic microwave background fluctuations, the sources extend far beyond recombination. Indeed they are only suppressed by the slow decay of the radiation density over the matter density. The time dependence of the sources is highlighted in Figure \ref{fig1}, showing the magnetic field at different redshifts multiplied by a factor of $(1+z)^2$ to cancel the decay of the magnetic field due to the expansion of the Universe. The curves overlap in the absence of sources. We find that magnetic field sources are still present after recombination, vanishing only after $1+z\approx 10$.

\begin{figure}[!htb]
	\includegraphics[width=8.5cm]{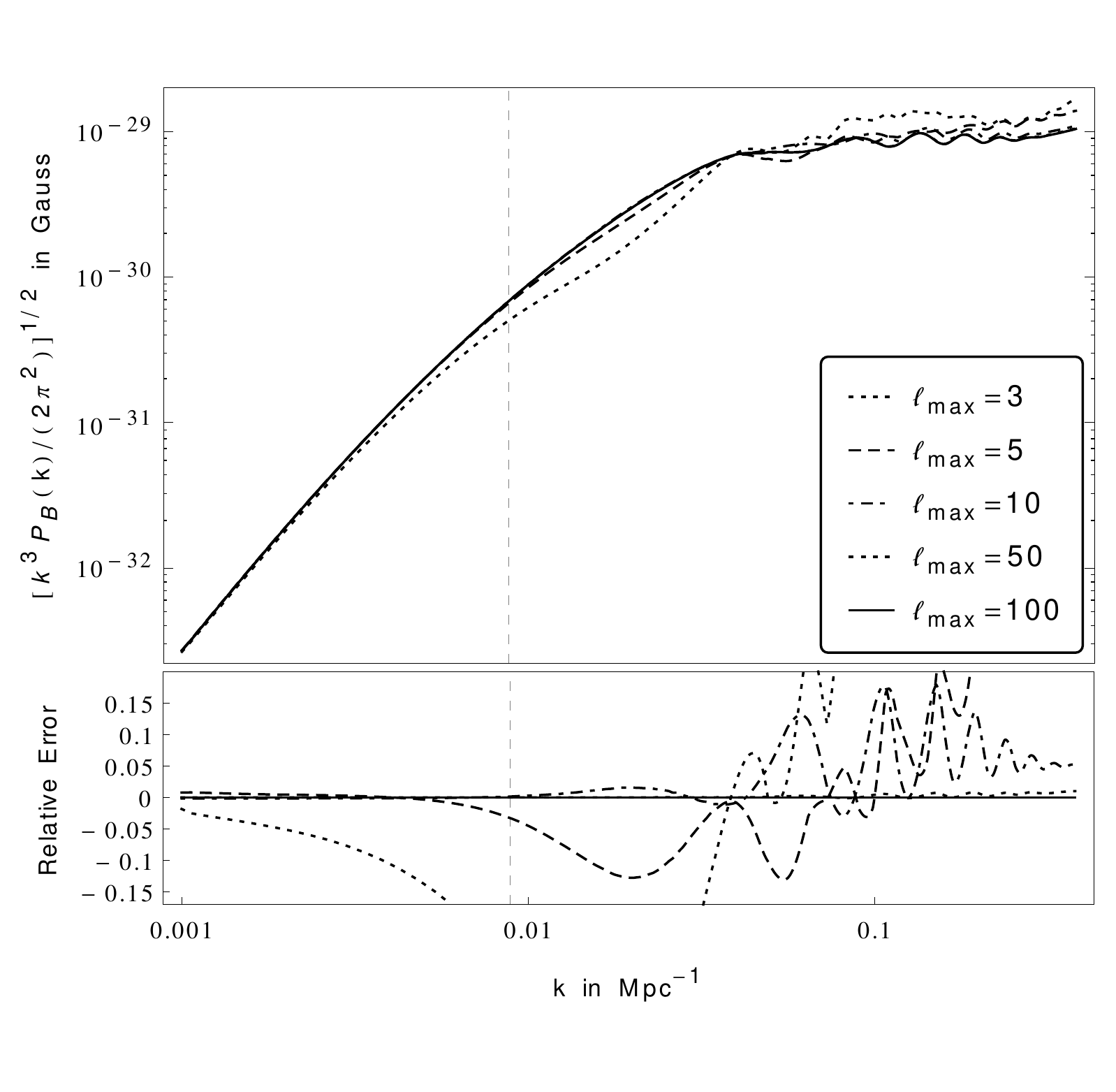}
	\caption{Stability of the magnetic field with respect to the angular resolution in \textsf{SONG}, represented by the maximum multipole $\ell_{\rm max}$ considered in the Boltzmann hierarchy. Large scales are not sensitive to this cut, whereas small scales need a cut at  $\ell_{\rm max} \approx 50$ to reach percent level accuracy.}
	\label{fig2}
\end{figure}

The free streaming of photons after recombination quickly generates small-scale oscillations characterised by large multipole moments, which have to been taken into account to correctly compute the magnetic field sources at low redshifts. While for the cosmic microwave background the photon hierarchy can typically be cut at $\ell_\text{max} \approx 10$, the magnetic field requires $\ell_\text{max} \approx 100$ at the scales of interest. For instance, if we consider sources at $z\simeq 100$, the number of multipoles needed in the Boltzmann hierarchy is of order $k_{\rm max} (\eta_{z=100}-\eta_{z_{\rm rec}}) \simeq k_{
\rm max} \times 1000 {\rm Mpc}$, so for a typical Fourier mode $k=0.1 {\rm Mpc}^{-1}$, we would need $\ell_{\rm max}\simeq 100$. Considering sources at lower redshift and larger Fourier modes in principle needs even more multipoles in the hierarchy. However the intrinsic magnetic field is only sourced by the multipoles entering Eq. \eqref{GenerationE}, that is the velocity $\ell=1$ and anisotropic stress $\ell=2$, simplifying the problem as we only need to compute the backreaction of the high multipoles on the low multipoles accurately. In practice we find that $\ell_{\rm max}\simeq 100$ is sufficient for the magnetic field on all analysed scales, while it is not sufficient for the the full photon hierarchy at the same scales. This complication was missed in the previous studies (\SIT~and \FPM), where only a few multipoles were considered, and leads to corrections larger than $10\%$ on a large range of scales as demonstrated in Figure \ref{fig2}.

The magnetic field is growing approximately by a power law on scales larger than $k_{\rm eq}$, the Fourier mode entering the Hubble radius at radiation/matter equivalence, indicated as a vertical dashed line in Figure \ref{fig2}. On smaller scales we find rich features including several oscillations imprinted from the interactions between the baryon and photon fluids. Finally, at the very small scales these oscillations are damped due to Silk damping.

\begin{figure}[!htb]
	\includegraphics[width=8.5cm]{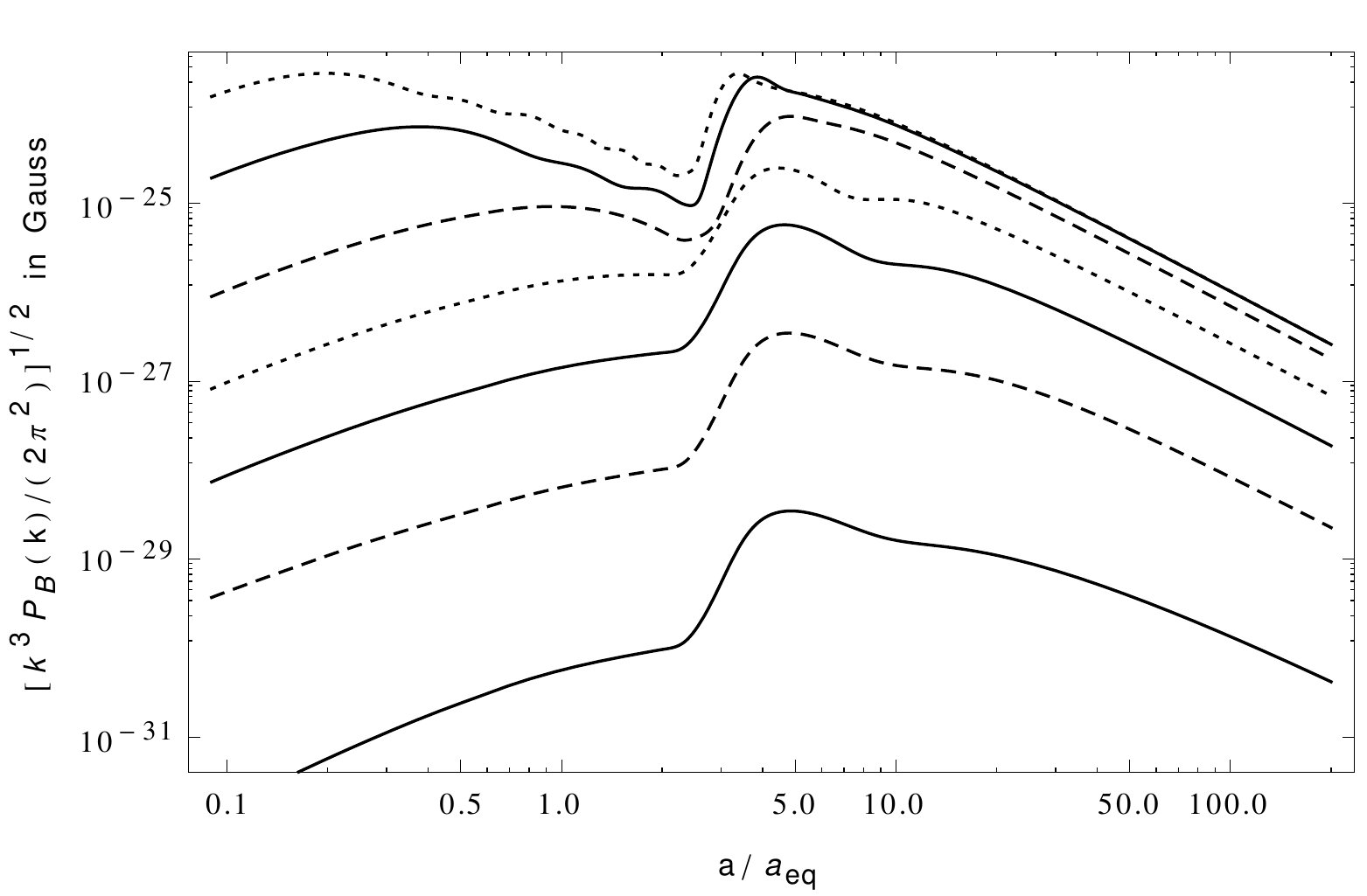}
	\caption{The time evolution of the magnetic field power spectrum for several k-modes. Solid lines are from bottom to top $k/k_{\text{eq}}= 0.1 , 1 , 10$, dashed lines are $k/k_{\text{eq}}= 0.4 , 4$ and dotted lines $k/k_{\text{eq}}= 2 , 20$. All modes grow at a constant rate while super-horizon and begin to decay once they enter the horizon. The peak is due to the generation of magnetic fields during recombination, and subsequently the evolution reverts back to constant decay once the photon density is negligible.}
	\label{fig3}
\end{figure}

In Figure \ref{fig3}, we show the temporal evolution of the magnetic field for various values of k. The magnetic field grows on super-horizon scales and starts to decay once it enters the horizon. 
Recombination creates a sharp bump in the magnetic field after which the magnetic field is still being sourced, combating the decay due to the expansion of the Universe until the amount of residual radiation is too small and the magnetic fields decay once more adiabatically as $1/a^2$.

\section{Comparison with analytical results and previous literature}

There has been a tension in the literature concerning the super-horizon limit at early times. In \SIT~it is found that $Q(k,\eta) \propto k^{3.5}$ while \FPM~finds a $k^4$ dependence thus questioning the validity of these results. We confirm the $k^{3.5}$ behaviour found in \SIT~numerically and provide hereafter for the first time an analytical argument explaining this super-horizon limit.

At second-order, the contributions to the electric field $E_i$ are of the form $\partial_i(A B)$, $A \partial_i B$ or $(\partial_i \partial_j A)(\partial^j B)$, where $A$ and $B$ are general first-order perturbations (see e.g. Eqs. (4-5) of Ref. \cite{Fenu2010}). From Eq.~(\ref{eq:Bevolution}), it can be seen that the former do not contribute to the magnetic field. Let us focus first on the terms of the type $A \partial_i B$. The symmetric convolution kernel ${\cal T}_B^i \equiv {\cal T}_B^+ e_+^i + {\cal T}_B^- e_-^i$ (related to its definition \eqref{DefBeta} in the polarization basis) associated with this type of term is of the form
\be\label{Bipropto}
{\cal T}_B^i \propto \varepsilon^{ijk}k_{1,j}k_{2,k}[{\cal T}_a(k_1){\cal T}_b(k_2)-{\cal T}_b(k_1){\cal T}_a(k_2)]\,,
\ee
with ${\cal T}_a(k)$ and ${\cal T}_b(k)$ being the linear transfer functions of $A$ and $B$ [e.g. $A(\gr{k},\eta)=T_a(k,\eta) \Phi_{\rm in}(\gr{k})$]. In \FPM, the limit in which all modes are super-Hubble has been considered (that is $k,k_1,k_2 \ll {\cal H}$  with ${\cal H}\equiv \partial_\eta \ln a$) and it has been shown that it leads to a $k^4$ dependence for $Q(k)$.

However, this is only a subpart of the convolution and we must also consider contributions where only $k$ is super-Hubble but $k \ll k_1 \approx k_2$. We find that these contributions dominate the convolution integral as they lead to a $k^{3.5}$ dependence for $Q(k,\eta)$. To see this, let us expand Eq.~\eqref{Bipropto} to leading order in $k$ using that $\gr{k}_2=\gr{k}-\gr{k}_1$:
\bea 
{\cal T}_B^i  &\propto&\varepsilon^{ijk}k_{1,j}k_k \left[{\cal T}_a(k_1){\cal T}_b(k_1)-{\cal T}_b(k_1){\cal T}_a(k_1)\right]  \\
&&+\varepsilon^{ijk}k_{1,j}k_k \left[{\cal T}_a(k_1)\frac{\dd {\cal T}_b(k_1)}{\dd k_1}-{\cal T}_b(k_1)\frac{\dd {\cal T}_a(k_1)}{\dd k_1}\right]k\,.\nonumber
\eea
The first line corresponds to a contribution linear in $k$ which vanishes while the second line gives the leading $k^2$ contribution. A similar reasoning with the terms of type $(\partial_i \partial_j A)(\partial^j B)$ leads to the same conclusion as it brings only an extra factor $\gr{k}_1 \cdot \gr{k}_2$. Hence, from Eq. \eqref{eq:PowerSpectrum} we deduce that the leading contribution to $P(k,\eta)$ scales as $k^4$, implying that $Q(k,\eta)$ scales as $k^{3.5}$. 

The analytical argument presented in \FPM~considers only the subleading contributions to the convolution integral when $k_1$ and $k_2$ are super-horizon, which do have a $k^4$ scaling.
In fact, in the numerical analysis in \FPM, the $q$-integration is cut at insufficient low values (a fixed multiple of the $k$ considered) restricting the integration to the super-horizon limit and therefore consistently finding the $k^4$ limit both analytically and numerically. This problem only affects the times before recombination when the horizon is small; the analysis of the magnetic field at present time in \FPM~is not affected by this cut since all modes are sub-Hubble at late times. Note also that the analytical analysis of Ref. \cite{Nalson2013} is also restricted to super-Hubble modes and they consistently find a $k^4$ dependence for $Q(k,\eta)$.

\section*{Conclusion}

We confirm that today's value of the averaged magnetic field generated from baryon-photon interactions is or order $2\times10^{-29}{\rm G}$ at a few Mpc, which is consistent with the values reported in the literature\footnote{Note that the results in \SIT~are not evaluated at $z=0$ but at $z=1100$ and they should be rescaled by an adiabatic decay to be expressed in terms of a magnetic field today; this point has been overlooked in \S 4.4.2 of~\cite{Widrow2011}, where the magnetic field was overestimated by a factor $\sim10^6$.}. We confirm the large scale limit of $k^{3.5}$ found in Refs.~\cite{Ichiki2007,Saga2015}, clarifying the tension in the past literature. At the small scales, we find that earlier numerical computations did not have enough angular resolution as they cut the Boltzmann hierarchy at insufficient values $\ell_{\rm max}$, not capturing the features left by the baryon-photons interactions correctly.
We conclude that Einstein-Boltzmann codes cannot be used to extend the power spectrum estimation at scales much smaller than $1 \,{\rm Mpc}$.
Contributions at low redshift are an important source for the magnetic field and at small scales they require both a large number of multipoles as well as beyond-second-order corrections. Since the fully relativistic theory is too challenging beyond the second order, a correct treatment of higher-order effects for small scales should be performed by using a perturbative expansion of the Newtonian theory on an expanding universe.

{\it Acknowledgements:} the authors would like to thank R. Maartens for his detailed comments on earlier versions of this article.

\bibliography{BiblioBSong}


\end{document}